# Type-II pumping beyond resonance principle: From energetic to geometric rules

B. Q. Song[1,2], J. D. H. Smith[1,3], Y. X. Yao[1,2], J. Wang[1,2]

[1]Ames National Laboratory, Iowa State University, Ames, Iowa 50011, USA
[2]Department of Physics and Astronomy, Iowa State University, Ames, Iowa 50011, USA
[3]Department of Mathematics, Iowa State University, Ames, Iowa 50011, USA

**Abstract**

Conventionally, pumping relies on energetic resonance: energy quanta $\hbar\omega$ matches the gap $\Delta$. Under linear approximation, this is known as the Fermi golden rule (FGR). However, this principle becomes challenging to apply in the "0/0" limit, where $\omega, \Delta \to 0$ simultaneously. In "0/0" scenarios, such as topological phase transition (TPT), a type-II pumping, *geometric pumping* (GP), is recognized subject to *geometric rule*s, distinguished from type-I dictated by FGR. Type-I features an "arrow of energy", sending particles higher in energy, reflected by FGR's dependence on Fermi distribution $f_v - f_c$ (probabilities of valence and conduction bands). While GP is non-directional, its probability relies on $f_v + f_c - 2f_v f_c$ instead, a key signature for detection. In this work, we address: (1) the concept of GP; (2) its features of fractionality, irreversibility, and dependence on TPT; (3) experimental detection with ultra-fast spectrum in coherent phonon driving of ZrTe$_5$.

**Energetic pumping & Fermi golden rule.** A basic principle for quantum transition is energy matching[1-3]: the driving frequency $\hbar\omega$ (photon, phonon, etc.) needs to be equal to the energy difference $\Delta$ (Fig. 1a)[4-7]. Thus, the energy ratio $\Gamma = \frac{\hbar\omega}{\Delta}$ is a useful characteristic, such as in Landau-Zener formula[8], the transition probability takes a form of $p \propto e^{-1/\Gamma}$. This is our familiar pumping, dictated by an energy ratio $\Gamma$, which can be called "energetic" or "type-I" pumping.

The Fermi golden rule (FGR)[1-3] is a special case derived under linear perturbation. In a two-band scenario, it is expressed by

$$p_E(\mu, T, \omega) = f_{c,v} \cdot p_E(\omega), (1)$$

where $p_E(\mu, T, \omega)$ is the $v \to c$ transition probability at chemical potential $\mu$ and temperature $T$ (label $k$ is ignored). $f_{c,v} \coloneqq f_v - f_c$ is Fermi distribution difference ($c, v$ refer to conduction and valence bands). $p_E(\omega)$ is the pumping rate at $T = 0$.

$$p_E(\omega) \propto |V_{c,v}(k, \omega)|^2 \cdot \delta\left(\omega - (\omega_c(k) - \omega_v(k))\right), (2)$$

where $V_{c,v}(k, \omega)$ is the perturbation potential. Equation. (1) is "energetic", because $p_E(\omega)$ clearly relies on $\hbar\omega$ and the gap. The energetic principle is formulated with a $\delta$-function: $\delta(\omega - (\omega_c - \omega_v))$, which is substantial at resonance $\Gamma = \frac{\hbar\omega}{\Delta} = \frac{\omega}{\omega_c - \omega_v} \sim 1$.

Another important feature of $p_E$ is the "arrow of energy", i.e., pumping is directed to the higher energy (Fig. 1a). If we reverse the population, i.e., conduction is filled, and valence is empty, pumping will send electrons even higher. The roles for the two bands distinguished by the energy arrow can be seen from the sign reversal of $f_{c,v}$ under exchange $c \leftrightarrow v$.

$$f_{c,v} = -f_{v,c} \quad (3)$$

Thus, the conventional pumping possesses two main features: (i) the probability contains an energy ratio $\Gamma$; (ii) particles tend to be pumped to the higher energy, termed *directionality*.

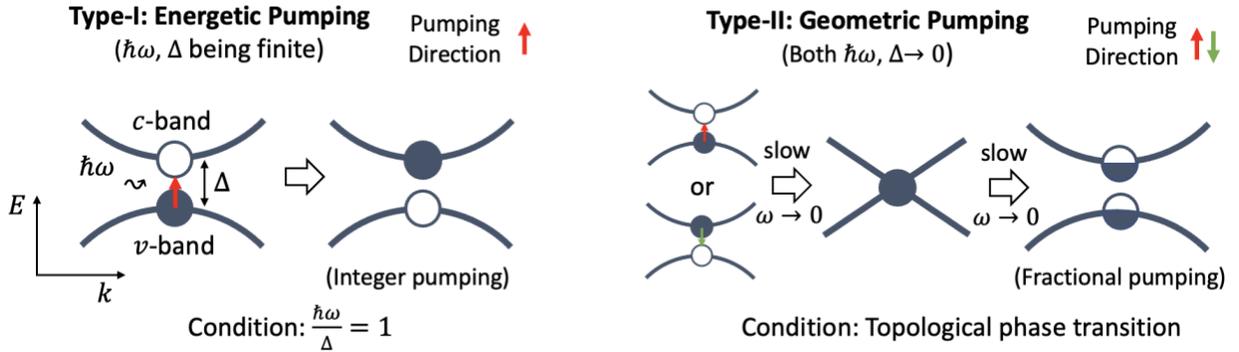

Figure. 1. Two types of pumping between bands. (a) Energetic pumping is excited by finite $\hbar\omega$ that matches finite gap $\Delta$. Under reversing population (lower panel) of particles (solid) and holes (hollow), pumping will send particles to higher bands, i.e., energetic pumping is directional. (b) Geometric pumping (GP) arises from simultaneous $\hbar\omega \to 0$ and $\Delta \to 0$, making $\Gamma = \hbar\omega/\Delta$ ill-defined, e.g., adiabatic band evolution causing gap to close/reopen. Reversing the particle-hole populations will end up with the same final state: both $v$- and $c$-bands are half filled, equivalent to pumping ½ particle from $c$-band to $v$-band. Thus, GP happens in both directions.

Type-I appears to cover every possible pumping scenario; however, it implicitly assumes that $\Delta$ should remain relatively constant during a driving cycle. This allows for a well-defined value of $\Gamma = \frac{\hbar\omega}{\Delta}$ and an identifiable resonance $\Gamma = 1$. Conversely, if $\Delta$ varies significantly over time, such as when the gap closing happens $\Delta = 0$, $\Gamma$ might be ill-defined. Thus, the type-I leaves some "shadow" where it does not fully apply.

One such scenario is the "0/0" limit, which concerns adiabatic driving ($\omega \to 0$) that slowly closes up the band gap $\Delta = (\omega_c(k_0) - \omega_v(k_0)) \to 0$. In this situation, $p_E(\omega)$ in equation (2) becomes divergent. Notably, this divergence does *not* stem from the "infinite spike" in the $\delta$-function (which is typically replaced with a finite Lorentz form), but rather from the undetermined relative rates at which $\omega$ and $\omega_c - \omega_v$ approach to zero. Physically, it relates to the breakdown of linear truncation near the point of gap closing. The two "zero energies" are encountered in many scenarios, such as transport[9-13], quantum criticality (both $k_B T$ and quantum critical frequency $\hbar\omega_c \to 0$)[14,15].

In this paper, we examine a particular 0/0 in bands at topological phase transition (TPT) of bands[16-21], where $\omega$ refers to the slow driving of phonons, and gap closing $\Delta \to 0$ is necessary for TPT. We identify a novel pumping that is solely dependent on TPT, independent of any energy ratio $\Gamma$, thus different from type-I. The argument is supported by both numerical and analytic results. Its distinguishable features include *Fractionality* and *non-directionality*, which should be interpreted statistically. Therefore, in the shadow, where FGR fails and new pumping rises, we discover an entrance for geometry into quantum dynamics and statistical physics, adding to its previous merits in classifying static eigenstates.

## Results

**Type-II: Geometric pumping**. We propose a second type of pumping (Fig. 1b), termed *geometric pumping* (GP). Analogous to energetic pumping, which is governed by an energy ratio $\Gamma$, GP is defined as dictated by geometric parameters. Essentially, GP represents geometric observables in "shadows" where traditional energetic rules fail to apply, and the 0/0 limit is a plausible scenario.

The challenge for unveiling such a phenomenon lies, firstly, in the limit of 0/0, which diverges under standard perturbation results, such as equation (2). For non-perturbation techniques, they face limitations by various specifications, e.g., interaction$\to \infty$[15,22], setting gaps with experiments. These difficulties undermine our understanding about "what would exactly happen at 0/0". Secondly, although numerical simulations provide clues, such as correlation of observables with geometric parameters, a rigorous conceptual foundation still relies on analytic solutions which are elusive to find.

Here we examine GP in a time-dependent two-band model (one filled and one empty) driven by phonons, where TPT might happen and serve as the geometric variable. Our purpose is to show the probability $p_G \neq 0$ for particles being pumped to the upper band after numerous cycles near gap closing $k_0$, and furthermore demonstrate $p_G$ only depends on TPT.

$$H(k) = \begin{pmatrix} -\epsilon_0 - A_{ph} \cdot \sin(\omega t) - \cos(k) & -i\sin(k) \\ i\sin(k) & \epsilon_0 + A_{ph} \cdot \sin(\omega t) + \cos(k) \end{pmatrix} \quad (4)$$

The periodic potential $A_{ph} \cdot \sin(\omega t)$ describes band distortion by phonons, which might close and reopen the gap within the cycle. If gaps are constantly open, a sufficiently slow $\omega$ ($\frac{\hbar\omega}{\Delta} \ll 1$) always exist to guarantee adiabaticity, and all particles will remain in the lower band leading to $p_G = 0$. Therefore, $\Delta_{\min} = 0$ is the key to go beyond adiabaticity, making $p_G \neq 0$ possible. Note that, although gap closing $\Delta \to 0$ is readily achieved, $\omega$ must be finite in practice (otherwise, $\omega = 0$ will freeze the evolution). The actual meaning of $\omega \to 0$ is that $\hbar\omega \ll$ average gap $\bar{\Delta}$ (average over an entire phonon period), as always assumed here.

We will first employ numerical methods to calculate $p_G$ in the presence and absence of TPT to demonstrate their correlation, suggesting $p_G$ is likely to be a type-II pumping. Then, we resort to analytic solutions, which endorse the numerical, to provide the conceptual ground for GP.

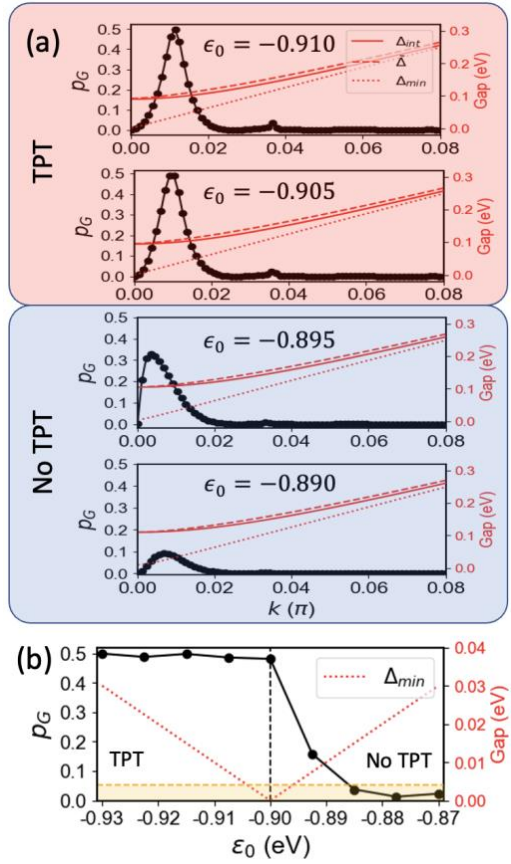

Figure 2. (a) Geometric pumping $p_G$ (black) in $k$-space; The pristine gap (without phonon) $\Delta_{int}$, average gap $\bar{\Delta}$ during an entire phonon period, and the minimum gap $\Delta_{min}$ at a local $k$. When $\epsilon_0 < -0.90$, $\Delta_{min} = 0$ at $k = 0$, which means phonons will close up the gap at $k = 0$ leading to TPT. If $\epsilon_0 > -0.90$, $\Delta_{min} > 0$ throughout BZ, and there is no TPT. (b) Maximum $p_G$ (peak heights like in (a)) sharply drops at the critical $\epsilon_0 = -0.90$ (dashed), while it is insensitive to energetic parameters $\Delta_{min}$ at $k = 0$ (red). Phonon energy (1 THz ~ 4 meV) is the orange shadow.

Numerical simulation is based on Trotter decomposition (Method)[23,24]. In this context, it means the following decomposition has error scales $(\Delta t)^2$.

$$e^{-\frac{i}{\hbar}\int_0^t H(\tau)d\tau} \approx e^{-\frac{i}{\hbar}H(t_N)\Delta t} \ldots e^{-\frac{i}{\hbar}H(t_2)\Delta t} e^{-\frac{i}{\hbar}H(t_1)\Delta t} \quad (5)$$

For each segment $\Delta t$, we may further perform an expansion,

$$e^{-\frac{i}{\hbar}H(t_j)\Delta t} = 1 + \frac{-i}{\hbar}H(t_j)\Delta t + \frac{1}{2!}\left(\frac{-i}{\hbar}H(t_j)\Delta t\right)^2 + \cdots. \quad (6)$$

The simulation result is presented in Fig. 2, which demonstrates $p_G$ is dependent on TPT, but insensitive energetic gaps. TPT is switched on (off) by $\epsilon_0 < -0.90$ ($> -0.90$) subject to a fixed $A_{ph} = 0.1$. Fig. 2a shows $p_G$'s distribution in BZ. Interestingly, maximum $p_G$ takes place in the vicinity of gap closing ($k_0 = 0$), rather than right on $k_0$. This is different from equation (2) which suggests maximum $p_G$ at minimum $\bar{\Delta}$ given $\hbar\omega \ll \bar{\Delta}$. Another distinct feature is noticed in $\epsilon_0$-space. The gaps are only minorly affected by $\epsilon_0$ (red in Fig. 2a), while $p_G$ is significantly suppressed by $\epsilon_0 < -0.90$, when TPT is turned off. A high-resolution scan for $\epsilon_0$ (Fig. 2b) exhibits a "step" at the critical point. All these indicate the observed pumping is *not* a phenomenon describable by energetic principles. Instead, it relies on the presence or absence of TPT.

Next, we show analytic $p_G$ to reinforce numerical results. Basically, we project the band problem to a spin model, because if $k$ is conserved, a two-band model driven by phonons is equivalent to a spin 1/2 driven by a cyclic **B**-field. Accordingly, the band evolution is modelled by the evolution $\mathcal{U}$ for spin in $\mathbf{B}(t)$. Since $p_G$ is the stable population after many cycles, we solve

$$p_n = \frac{1}{n}\sum_{j=1}^{n} p_j = \frac{1}{n}\sum_{j=1}^{n} |\langle n_1|\mathcal{U}^j|\varphi(t=0)\rangle|^2 , (7)$$

where $|\varphi(t=0)\rangle$ stands for the ground state, and $|n_i\rangle$ stands for spin eigenstates ($i = 0,1$ means lower/upper levels), $p_G$ is the limit

$$p_G := \lim_{n\to\infty} p_n . (8)$$

When the spin problem is solved, we project it back to bands, giving below the analytic result (equation (9)). It endorses the numerical finding: $p_G$ depends on TPT (Fig. 2) rather than energy ratio $\Gamma$, confirming the concept of GP.

$$\begin{cases} p_G = \frac{1}{2}, \text{TPT} \\ p_G = 0, \text{No TPT} \end{cases} (9)$$

The analytic solution addresses challenges such as when existing methods assume either eigenstates are fixed[4-6], or that occupancies (e.g., of ground states) are fixed[25,26]. The procedure of solving analytic $p_G$ based on quantum Liouville's theorem[21] and related backgrounds [27,28] are introduced later in method. Here, we first focus on physical questions. (1) Fractionality and non-directionality of the pumping. (2) Interpretation of $p_G$ (i.e., how $p_G$ is linked to observables). (3) The "dephasing" associated with GP. (4) The signatures for GP distinctive from FGR for experimental detection.

**Fractionality and non-directionality.** Fractionality means GP (between two bands) cannot exceed ½ at a local $k$, as shown by numerical (Fig. 2) and analytic results equation (9). This is a key signature for type-II pumping, distinguishable from direct pumping by phonons or other quanta. Because, in those cases, pumping probability may → 1, i.e., all particles will be pumped if there are sufficient phonons to stimulate. In Fig. 3a, we show the $p_G$'s dependence on $A_{ph}$, which reflects the amount of incoming phonons, and find ½ is indeed the "ceiling". On the other hand, an increased $A_{ph}$ may make the peak "fatter", which reconciles with the conventional belief under a coarse averaging. For the super-adiabatic process $\omega \to 0$, the peak will get super-sharp, approaching to the dual values ½ and 0, as suggested by the analytic results equation (9).

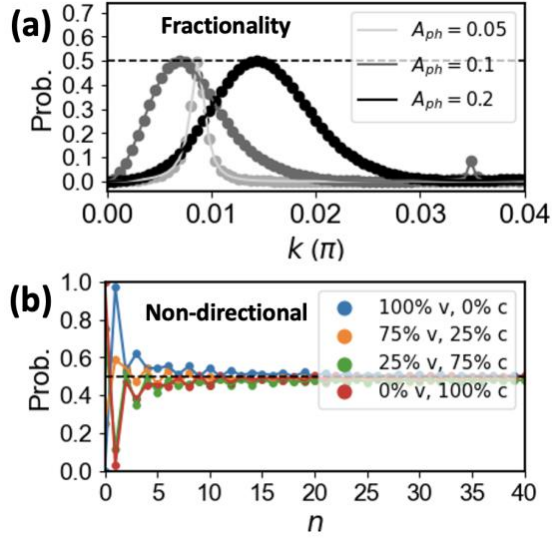

*Figure 3. (a) Probability of pumping under different phonon amplitude. (b) Different initial states converge to the same final distribution. The "v" and "c" refer to valence and conduction bands.*

Another signature for GP is "non-directionality": pumping lacks an "arrow of energy". We examine initial states (Fig. 3b) of different weights in valence and conduction bands and find they converge to the same destination of ½. The "non-directionality" means the GP is more like a "mixing": if it starts with the ground state (blue in Fig. 3b), GP tends to pump (at a local $k$) half of the electron to the conduction band; if it starts with the excited conduction state (red in Fig. 3b), GP will send ½ particle back. GP will disappear if two touching bands are both filled or both empty.

**Ensemble interpretation of $p_G$.** In equation (7), $n$ refers to the cycle number, acting roles as time. Thus, $p_n$ means the average over all the past $n$ cycles. This applies to $\tau_{measure} \gg$ cycle period $\tau_{cycle}$, i.e., the system has traversed numerous cycles $\mathcal{U}^n$. (Otherwise, if $\tau_{measure} \lesssim \tau_{cycle}$, equation (7) should not contain $\frac{1}{n}\Sigma$.) Equation (7) is a standard form of observable in statistical mechanics[29,30]. It can be called *long-time interpretation* (Fig. 4a), and $p_G \coloneqq p_\infty$ describes stable situations such as in equilibrium.

Alternatively, there is a second interpretation for equation (7), the *ensemble interpretation*. Consider $n$ identical systems (Fig. 4b), which start from the same initial state, say the ground state (GS), but which join the evolution at different moments, i.e., each entry passes distinct times $t_j$ in $e^{-iHt_j}$. Observables are still obtained with equation (7), while $p_j$ is interpreted as the $j^{th}$ entry in the ensemble, instead of the single system at the $j^{th}$ moment. Accordingly, $p_n$ is the average of $n$-systems at an instant (Fig. 4b), rather than the long-time average of the single system (Fig. 4a). Simply speaking, the long-time average is now re-interpreted as an ensemble average. We call $p_n$ the *ensemble probability* for the $n$-system, and $p_G$ (equation (8)) originally referring to a long-time limit, represents a large-size the ensemble.

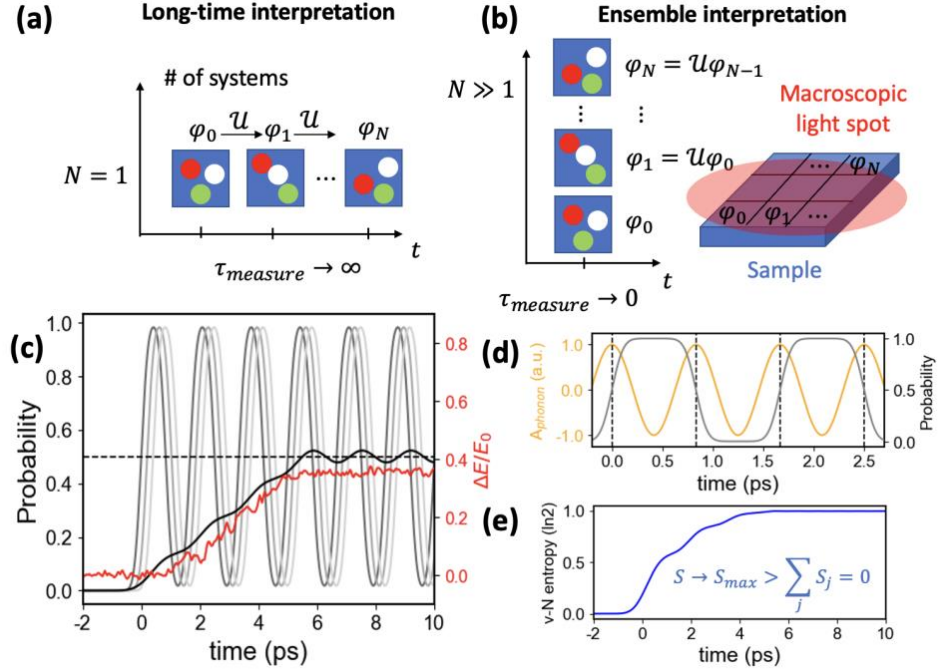

*Figure 4. (a)(b) Two interpretations. (c) The grey lines are $p_j(t)$ (sampling $j = 1,2,3$), with small $\Delta t$ mismatch. Each $p_j(t)$ is coherent, while $\sum_j p_j(t)$ displays a plateau (black line), saturated at $p_G = \frac{1}{2}$, consistent with observations of smooth inter-band pumping in ZrTe5 (red dots, pumping charge measured by the transition rate change compared with GS, $\Delta E \propto Q_{pump}$)[19]. (d) $p_1(t)$ (grey) and its corresponding phonon (orange). TPT is assumed once per cycle (otherwise, effective $\tau_{cycle}$ should be shorter), occurring at the maximum instantaneous amplitude $A_{phonon}(t)$. $p_1(t)$ is calculated by $\mathcal{U}^n$ with $\Theta = \pi$, where $n$ is the times of TPT the system has traversed; between two TPT $p_1(t)$ should remain, as inter-band decay is slow when the gap is present (i.e., $p_1(t)$ changes mainly at TPT, marked by dashed lines). (e) The entropy associated with evolution in (c).*

Ensemble interpretation works for a measurement of sharp time resolution $\tau_{measure} < \tau_{cycle}$ but a coarse spatial one. Several ultra-fast[12,13,19,31,32] and terahertz spectroscopy responses[33-36] are of this type. Basically, this technique shines laser on the sample and measures the response in reflectivity, transition rates, etc., with time resolution ~ 10s fs, much smaller than the phonon cycle $\tau_{cycle}$ ~ 1 ps; on the other hand, the laser spot could be in mm levels (Fig. 4b), and each microscopic region in the light spot corresponds to a $p_j$ in the ensemble. Next, we apply $p_G$ to ZrTe5[19] to address the pumping and saturation observed therein. The data of pump-probe experiment (Fig.2a in Ref. [19]) is re-plotted (red in Fig. 4c). The change of transition rate $\Delta E$ provides real-time monitoring of inter-band pumping charges $\Delta Q_{pump} \propto \Delta E$. (Experimental techniques are further introduced in Sec. 1 of supporting information, SI).

Two key observations. (1) Pumping (red in Fig. 4c) happens in a sub-gap regime: $\hbar\omega$~4 meV and average gap $\bar{\Delta}$~ 40-100 meV. (2) The excited $A_{1g}$ phonon lasts for $\tau_{phonon}$~100s ps[19], much longer than the duration of pumping ~5 ps. Both seem abnormal to conventional mechanisms. Type-II can address them. Firstly, $p_G$ is caused by TPT, such as the transition between strong and weak topological phases in ZrTe$_5$[37,38]. Phonons facilitate gap closing and TPT rather than "directly" exciting particles, explaining why pumping occurs even when $\hbar\omega \ll \bar{\Delta}$. Secondly, $p_G$ is saturated when $\mathcal{U}^{n+1}$ causes no difference from $\mathcal{U}^n$. Physically, this signifies a "dynamic balance": the pumping and de-pumping in $n$-systems equalize, without necessitating the termination of TPT. This explains why pumping disappears earlier than the phonon does.

The "plateau" in Fig. 4c corresponds to $p_G = \frac{1}{2}$ suggested by equation (9). Physically, $p_G = \frac{1}{2}$ implies that after a few phonon cycles, the system will reach a stable distribution: the conduction and valence bands are half-filled near the gap closing $k_0$. The saturation time $\tau_{ergodic}$ of reaching the "plateau" is neither instant nor infinite. Under the ensemble interpretation, it is the process of entries in the ensemble increasing from 1 to ∞. Physically, one small region in the light spot starts to vibrate, one member will be added to the ensemble. Thus, $\tau_{ergodic}$ corresponds to the time for everywhere within the laser spot building up phonons. The process of reaching the "plateau" could be simulated with *time-dependent ensemble.*

$$\mathcal{O}_\mathcal{E}(t) = \frac{1}{n}\sum_{j=1}^{n} \mathcal{O}_j(t) \quad (10)$$

where $\mathcal{O}_\mathcal{E}$ is an observable in ensemble $\mathcal{E}$ of $n$ systems (in the case, $\mathcal{O}_\mathcal{E}(t) \sim p_n(t)$, $\mathcal{O}_j(t) \sim p_j(t)$). $\mathcal{O}_j(t)$ is associated with the $j^{th}$ system at $t$. Equation (10) is generalization of equation (7) by adding time dependence; accordingly, the static ensemble $\{\mathcal{O}_j\}$ is generalized to a set $\mathcal{E}(t) \coloneqq \{\mathcal{O}_j(t)\}$, called a *time-dependent ensemble*. The ensemble observable $\mathcal{O}_\mathcal{E}(t)$, the average of the $n$ systems at $t$, changes with time. As such, equation (10) can describe the process of achieving a stable state of $\mathcal{E}(t)$, which means a dynamic balance of $\dot{\mathcal{O}}_\mathcal{E}(t) = 0$ (however, $\dot{\mathcal{O}}_j(t) \neq 0$).

$\{\mathcal{O}_j(t)\}$ stand for a series of systems that mismatch in evolution times $p_{j+l}(t) = p_j(t - l\Delta t)$. Physically, that means different small regions mismatching in starting times. Then, equation (10) becomes

$$p_\mathcal{E}(t) = \frac{1}{n}\sum_{j}^{n} p_1(t - j\Delta t) \quad (11)$$

$p_1(t)$ (the earliest in $\mathcal{E}(t)$) could be evaluated by $\mathcal{U}^n$ (method). The von Neumann entropy $S(\hat{\rho}) = -\text{tr}(\hat{\rho}\ln\hat{\rho})$ is plotted in Fig. 4e. In the dephasing process, each system performs a unitary reversible evolution, i.e., $S_j = 0$, (grey lines in Fig. 4c) while the total $S$ tends to $\ln 2$. This is due to concave condition: $S(\sum_j \lambda_j \hat{\rho}_j) \geq \sum_j \lambda_j S(\hat{\rho}_j)$. Thus, GP's dephasing arises internally, rather than by coupling with an external heat bath.

**Geometric rules & Experiment signatures.** GP rationalizes how a sub-gap driving possibly leads to pumping smooth over time (rather than a "spike" at the gap matching $\hbar\omega$[39]), and saturated before the decay of phonons. To reinforce the argument, we next derive the GP's signatures against FGR that can be tested with dependences of temperature, laser fluence, etc.

*Table I: GP occurs when only one of v-band and c-band is occupied (Occ.), which correspond to probabilities $(1 - f_v) \cdot f_c$ and $f_v \cdot (1 - f_c)$. The finite-temperature factor $g_{c,v}$ is calculated by $0 \cdot f_v \cdot f_c + \frac{1}{2}(1 - f_v) \cdot f_c + \frac{1}{2}f_v \cdot (1 - f_c) + 0 \cdot (1 - f_v) \cdot (1 - f_c) = f_v + f_c - 2f_v \cdot f_c$.*

| $v$-band | $c$-band | GP  | Prob.                    | $p_G$ |
|----------|----------|-----|--------------------------|-------|
| Occ.     | Occ.     | No  | $f_v \cdot f_c$          | 0     |
| Occ.     | Emp.     | Yes | $f_v \cdot (1 - f_c)$    | ½     |
| Emp.     | Occ.     | Yes | $(1 - f_v) \cdot f_c$    | ½     |
| Emp.     | Emp.     | No  | $(1 - f_v) \cdot (1 - f_c)$ | 0  |

At $T = 0$, $p_G$ is given by equation (9); at finite $T$, $p_G$ is evaluated by Table I, which we call *geometric rule*

$$p_G(\mu, T, \Delta v) = g_{c,v} \cdot p_G(\Delta v), (12)$$

in analog with FGR equation (1). $p_G(\Delta v)$ is a compact form of equation (9): $p_G(\Delta v) = \frac{1}{2}$ with $\Delta v = 1$, and $p_G(\Delta v) = 0$ with $\Delta v = 0$. ($\Delta v$ is the change of topological index) $p_G(\Delta v)$ corresponds to $p_E(\omega)$: energetic parameter $\hbar\omega$ is replaced by geometric index $\Delta v$. By analog with $f_{c,v}(k)$ in energetic pumping, $g_{c,v}(k)$ is defined as

$$g_{c,v} := f_v + f_c - 2f_v \cdot f_c . (13)$$

Properties of geometric rule. First, it will preserve the sign under band exchange, in contrast with sign reversal in equation (3). This indicates that as "0/0" is approached, arow of energy disappears. Intuitively, it can be rationalized by the 0/0 limit disregarding the differences between higher- and lower-energy bands.

$$g_{v,c} = g_{c,v} (14)$$

Second, positive definite: $g_{v,c} = f_v + f_c - 2f_v \cdot f_c = f_v(1 - f_c) + f_c(1 - f_v)$. Since $0 \leq f_{c,v} \leq 1$,

$$0 \leq g_{v,c} \leq 1 (15)$$

in analog with $0 \leq |f_{v,c}| \leq 1$.

Third, geometric rule differs from FGR at band degeneracy: $E_v = E_c$, which leads to $f_v = f_c = f$. Then we have $f_{v,c} = f - f = 0$, while $g_{v,c} = f + f - 2f^2 = 2f(1 - f) > 0$ for $T > 0$. That means that while FGR demonstrates as a dip, $g_{v,c}$ usually demonstrates as a peak. Since degeneracy usually leads to TPT, GP should feature a peak at TPT.

Both $p_E(\mu, T, \omega)$ and $p_G(\mu, T, \Delta v)$ rely on $T$ and $\mu$, however, by means of $f_{v,c}$ and $g_{v,c}$, respectively. Detections should be with respect to $T$ and $\mu$. Note that $p_G(\mu, T, \Delta v)$ is non-energetic, Fourier analysis (response functions in $\omega$ space) should not be taken for granted, and $p_G(\mu, T, \Delta v)$ is not directly comparable with $p_E(\mu, T, \omega)$ against driving frequency $\omega$.

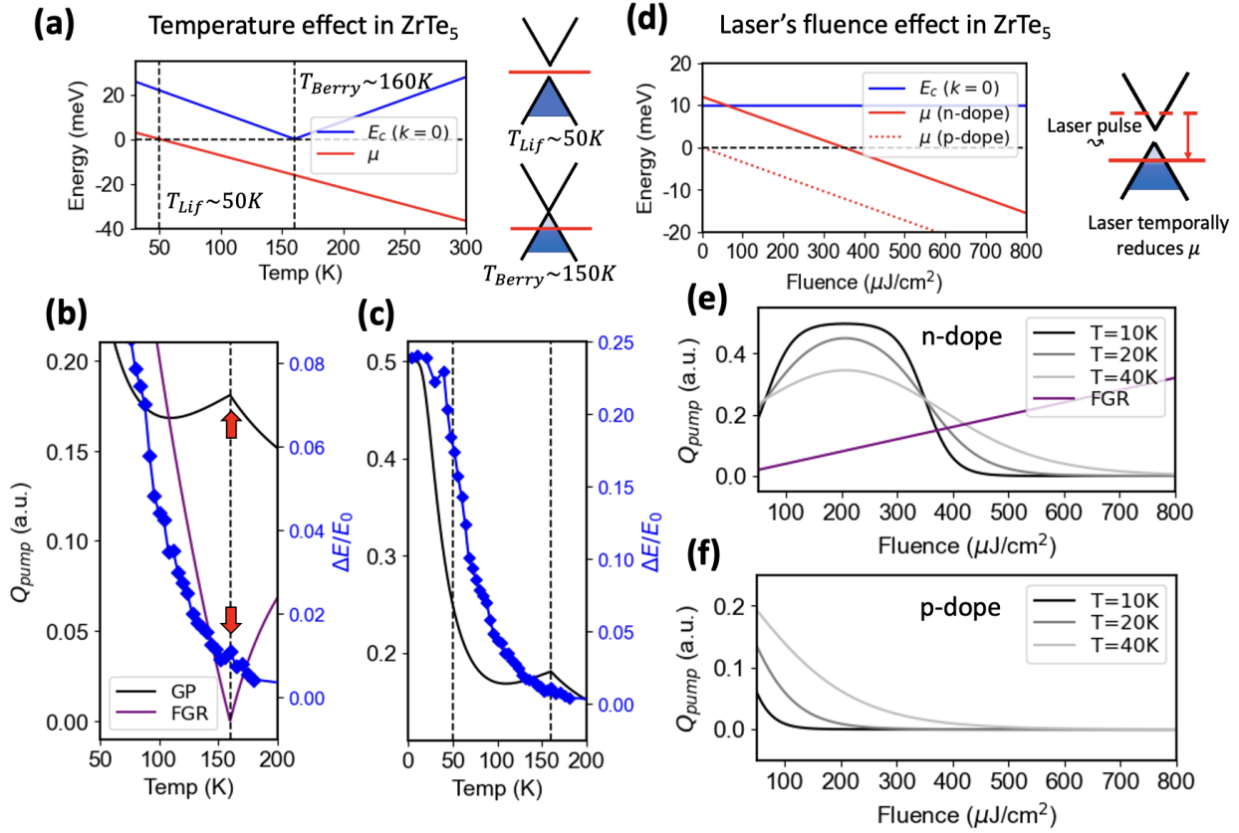

Figure 5. (a) A model of linear $T$-dependence. At $T_{Lif} \sim 50\ K$, gap is finite and $\mu = 0$. At $T_{Berry} \sim 160\ K$, gap is zero and $\mu < 0$. (b)(c) GP evaluated based on model as (a), compared with FGR and experiment[19]. (d) A model of linear fluence-dependence with an achievable fluence range in experiment. (e)(f) Fluence dependence simulated based on the linear model as (d).

A key distinction between type-I and type-II lies in their dependence on the Fermi distribution functions: the energetic rule $f_{c,v} := f_v - f_c$ for type-I and the geometric rule $g_{c,v} := f_v + f_c - 2 f_v \cdot f_c$ for type-II. To experimentally test this, one could investigate the differences between $f_{c,v}$ and $g_{c,v}$ in their temperature and fluence dependences.

Temperature affects chemical potential and band gaps. In ZrTe$_5$, it observes that the $n$-doping at low-$T$ will turn into $p$-doping at high-$T$ around 50 K, the so-called *Lifshitz transition*[37]. In addition, due to thermal change of lattice parameters, the band gap (~10s meV) is closed around 150~160 K, bordering a low-$T$ strong and high-$T$ weak topological phases[37,38]. Based on these facts, Fig. 5a shows a linear model that features: (i) at $T = 0$ gap ~ 10s meV, (ii) at 50 K $\mu = 0$, (iii) at 160 K, gap is zero. In Fig. 5b, we compare FGR (purple) with GP (black) within the model. GP features a peak at TPT, while FGR features a dip. As mentioned above, this is due to $f_c = f_v$, $f_{v,c} = 0$ at degeneracy but $g_{v,c} > 0$. The $Q_{pump}$ can be measured by transition change $\Delta E$ with the pump-probe technique (SI), which exhibits a cusp around 160 K, against FGR. Even with the super-coarse linear model, $g_{v,c}$ captures the experimental observations around 160 K (Fig. 5b), as well as in a bigger range (Fig. 5c). The low-$T$ enhancement (< 50 K) is due to the gap being larger and thermal spreading being weaker, which makes $f_v \to 1$, $f_c \to 0$, such that $g_{v,c} \to max = 1$.

To tune $\mu$, one can use a short laser pulse (e.g., 800 nm, ~50 fs) to temporally empty the occupied states near the Fermi level, i.e., effectively decrease $\mu$[40] (Sec. 2 of SI). For simplicity, we assume $\Delta\mu$ is linear with laser fluence and will not affect the band structure (Fig. 5d). In Fig. 5e, f, we show the fluence dependence of $Q_{pump}$ in $n$- and $p$-dope scenarios. The most distinguishable feature is an abnormal trend: $Q_{pump}$ decreases with laser pulse intensity. Such a negative response is counterintuitive. But remember GP is not directly by electric field, it is due to band evolution, repeated TPT (Sec. 2 of SI); electric field only "lights the fire" (like the role of thermal fluctuation in phase transition). On the other hand, FGR suggests fluence~$|V_{c,v}|^2$, and $Q_{pump}$ should linearly increase with fluence. This feature is robust with temperature (Fig. 5e, f). The $n$-doping and $p$-doping differ in their trends in the low fluence regime, which can be tested by examining samples of different doping levels. Such empirical evidence is also achievable by tuning biased voltages in a proper setting.

**Discussion.** Figuratively, GP (under ensemble interpretation) can be imagined as a formation of marching individuals, initially in harmony and everyone being on the same footing. Given *no* interference between individuals (so everyone follows their own rhythm), the tiny mismatch will accumulate, and after a certain while ones on their left and right feet get equal – a "pumping" from one foot to the other, and the formation's steps become random. This resembles the fact that geometric dephasing is due to self-propagation, rather than coupling with external heat bath.

On the other hand, there are further inquiries such as why is GP linked to TPT? Why does $p_G$ lead to a particular fractional value? Why is $p_G$ robust to model details? These may not be captured by everyday analog but must be understood with involved evaluations. This is why GP is more intriguing than "footsteps getting random".

*Table II: Comparison of energetic and geometric pumping in terms of rules to obey, variables, pumping direction, application scope, etc.*

|   | Type-I: Energetic $p_E$ | Type-II: Geometric $p_G$ |
|---|---|---|
| Rules | FGR: $f_{c,v} = f_v - f_c$ <br> $f_{c,v} = -f_{v,c}$ | Geom. Rule: $g_{c,v} = f_v + f_c - 2f_v \cdot f_c$ <br> $g_{v,c} = g_{c,v}$ |
| Variables | $\mu, T, \omega$ | $\mu, T, \Delta v$ (or other geom. Para.) |
| Formula | $p_E(\mu, T, \omega) = f_{c,v} \cdot p_E(\omega)$ | $p_G(\mu, T, \Delta v) = g_{c,v} \cdot p_G(\Delta v)$ |
| Directionality | Yes (to higher energy) | No (both) |
| Appl. Scope | Finite $\Delta$ | "0/0", $\Delta, \omega \to 0$ |

How does one pumping cross over to the other? If taking a coarse grain view of Fig. 2a, one finds the envelope is similar to a distribution described by a finite width $\delta$-function given by FGR. Thus, $p_E$ provides a description in larger ranges of $k$. On the other hand, if we get closer to $k_0$, we find the peak heights get saturated at ½, and merely depend on TPT. Thus, $p_G$ describes a finer picture for pumping, particularly close to $k_0$; in this small region, with finer resolutions, we should encounter a plateau as in Fig. 2a. Thus, the geometric rule is *not* denying energetic principles but is actually another facet of quantum rules when energy matching principle is hard to apply.

GP differs from trivial phonon-excited pumping in several ways: (1) GP is fractional, (2) GP is non-directional, (3) GP obeys geometric rules, (4) GP occurs off resonance with $\hbar\omega \ll \overline{\Delta}$. The pumping observed here is based on $\hbar\omega \ll \overline{\Delta}$ except for an infinitesimal period of gap closing in the cycle. This distinguishes it from adiabatic evolution, where $\hbar\omega \ll \overline{\Delta}$ constantly holds, and from Rabi oscillations, which result in pumping when $\hbar\omega \sim \overline{\Delta}$.

It remains unclear whether the classification by type-I and type-II is exhaustive. However, the present research suggests there are "shadows" or overlooked areas within type-I, and unknown pumping mechanism might emerge. For example, a strange aspect of type-II is fractionality (in the ideal limit, it is precisely a half particle). Fractionality is widely interested in physics[41-43], in condensed matter, represented by fractional quantum Hall effect (FQH)[44,45], fractional quantum anomalous Hall effect (FQAH)[46-49], fractional Chern insulator (FCI)[50,51], etc. However, these cases are based on strong interaction and transport phenomena. Here fractionality matters for (bulk) pumping phenomenon and is derived from non-interacting statistics, relying on quantum Liouville's theorem[21]. In view of the theorem's importance in statistical mechanics, we anticipate more from the interplay between topology and statistics.

Type-II extends topo-effects from surface[16,26,52] to bulk states, from (near) equilibrium[37,44] to non-equilibrium, expanding the techniques of measurement. In optical domains, techniques might include pump-probe spectroscopy[19,31 32], THz time-domain spectroscopy[33-36], ultra-fast X-ray[53], etc. These methods can conveniently induce TPT and provide real-time monitoring of electron or lattice motions during TPT. Authors demonstrate a specific example in the context of ultra-fast spectroscopy[19]. The experimental data are reasonably consistent with theory predictions (Fig. 4, 5). Notably, type-II pumping could decrease with increasing driving intensity, a striking feature that motivates further testing in broader systems using different techniques.

**Summary**. We recognize a type-II pumping mechanism, validated by numerical and analytic results. The characteristics of type-II pumping are detailed in Table II. This mechanism adheres to the geometric rule $g_{c,v} = f_v + f_c - 2f_v \cdot f_c$, which imparts distinguishable features. In ZrTe$_5$, notable signatures include: (i) Continuous pumping even though the band gap is significantly larger than the phonon energy; (ii) Enhanced pumping at TPT around 150-160 K, contrary to the dip predicted by FGR; (iii) Anomalous fluence dependence, where pumping decreases with increasing laser intensity.

## Methods.

In method, we address three questions. First, Trotter decomposition methods for numerical simulation of Fig.2. Second, the analytic expression for $p_G$. Third, the simulation for the time dependence of pumping curves in Fig. 4.

1. **Numerical method & Trotter decomposition.**

In general, Trotter decomposition is about breaking up the evolution into discrete time steps. Given a Hamiltonian $H$, one could employ the method numerically evaluate the evolution operator for a later time moment. The Trotter decomposition is originally a formula that approximates the exponential function of the sum of square matrices $A$ and $B$:

$$e^{(A+B)\delta} = e^{\delta A} \cdot e^{\delta B} + O(\delta^2). \quad (16)$$

In this context, it means approximating evolution operator

$$U(t,0) \approx \mathfrak{T} \prod_j^N e^{-iH(t_j)\Delta t} + O((\Delta t)^2) \quad (17)$$

where $\mathfrak{T}$ is time-ordered operator, and $\Delta t \coloneqq t/N$. In particular, it is about

$$U(t,0) = \left[1 + \frac{-i}{\hbar}H(t_N)\Delta t + \frac{(-i)^2}{2!\,\hbar^2}H^2(t_N)(\Delta t)^2 + \cdots\right]\left[1 + \frac{-i}{\hbar}H(t_{N-1})\Delta t \right.$$
$$\left. + \frac{(-i)^2}{2!\,\hbar^2}H^2(t_{N-1})(\Delta t)^2 + \cdots\right]\cdots\left[1 + \frac{-i}{\hbar}H(t_1)\Delta t + \frac{(-i)^2}{2!\,\hbar^2}H^2(t_N)(\Delta t)^2\right.$$
$$\left. + \cdots\right] \quad (18)$$

Therefore, it involves two truncations: (i) the time step resolution $N$, (ii) the truncation for polynomial expansion $H^n(t_j)$ of $e^{-iH(t_j)\Delta t}$. In principle, as Trotter decomposition scales as $O((\Delta t)^2)$, the linear truncation might be sufficient for (i). However, with the present model, it might lead to artifacts for (ii). For example, at $k = 0$, the linear approximation leads to

$$e^{-iH(t_j)\Delta t} \approx \left[1 + \frac{(-i)H(t_j)\Delta t}{\hbar}\right] = a + (-i)b\sigma_1 \quad (19)$$

where $a, b$ are certain real numbers and $\sigma_1$ is the Pauli matrix $\begin{pmatrix} 0 & 1 \\ 1 & 0 \end{pmatrix}$. Then, multiplication of two $e^{-iH(t_j)\Delta t}e^{-iH(t_{j+1})\Delta t}$ gives $(a + (-i)b\sigma_1)(a' + (-i)b'\sigma_1) = aa' - bb' - i(b + b')\sigma_1 = a'' - ib''\sigma_1$. Obviously, the diagonal is always real, and off-diagonal is always pure imaginary. However, this is an artifact due to an oversimplification. Therefore, it is important to maintain at least to the second order, which will provide a "tunnel" for the real and imaginary parts to mix. In addition, keeping the non-linear terms efficiently improve the unity of the evolution operator. The truncation of $e^{-iH(t_j)\Delta t}$ to finite orders of $H^n$ will make iteration deviate from unity.

For the time resolution, we choose $N = 20000$, and $t \sim 1$ ps. Thus, the time step of simulation is $\Delta t \sim 5.0 * 10^{-17}$ s. The test of convergence is shown in Sec. 3 of SI.

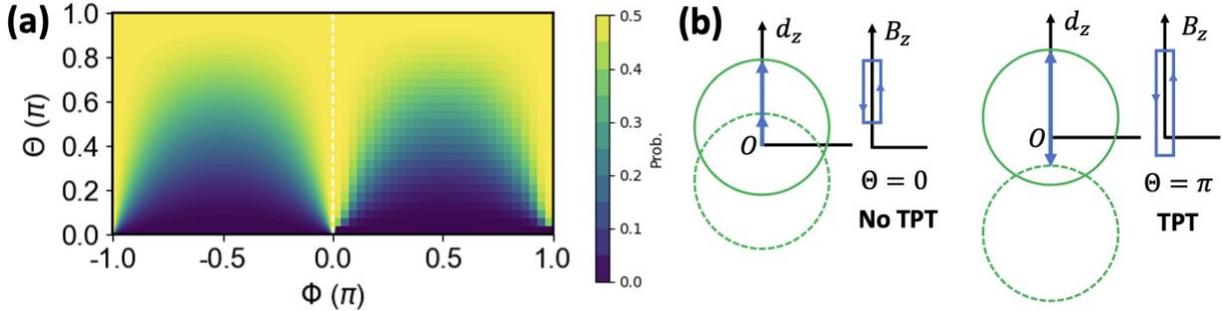

Figure 6. (a) Comparison of analytic (left) and numerical (right) solutions for $p_G$. For each pixel for the numerical solution, we have adopted $p_G \approx p_{n=100}$. (b) $\Theta = \pi$ (0) in spin model corresponds to TPT (no TPT) in band model. Each $k$ corresponds to a vector $d_i(k)$ which is like **B** vector. The green circle stands for 1D BZ; the dashed green indicates the band distortion. The topological state is characterized by the winding number of the circle with respect to the origin $O$.

2. **Analytic solutions for $p_G$**

We will derive analytic solutions that lead to equation (9) to prove $p_G$ is independent of energetic details, as suggested by numerical results (Fig. 2). Accurately, the analytic solution is for a model of spin ½ driven by cyclic B-fields, which represents the two-band model driven by phonon at a local $k$. The analytic solution is

$$p_G = \frac{1}{2} \frac{\sin^2\left(\frac{\Theta}{2}\right)}{1 - \cos^2\left(\frac{\Theta}{2}\right)\cos^2(\Phi)}, (20)$$

where $\Theta$ and $\Phi$ are parameters in spin models. The match of analytic Equation (20) and numerical results is self-explanatory (Fig. 6a). If project the spin back to the band model, we find $\Theta = 0$ corresponds to no TPT; $\Theta = \pi$ corresponds to TPT (Fig. 6b). Plugging the two values into equation (20), we find $p_G = 0$ and $\frac{1}{2}$, which is just equation (9). Next, we present derivation details in four steps.

(1) Based on independent $k$ (ignoring inelastic scattering), we reduce a two-band model to a spin ½ model at local $k$.

(2) Write down one-cycle evolution $\mathcal{U}$ in terms of corresponding (geometric) parameters in the spin model. In the new contexts, GP corresponds to the distribution of spin over the two spin eigenstates after numerical B-field cycles.

(3) Since $p_G$ merely relies on $\lim_{n\to\infty} \mathcal{U}^n$, one may skip the dynamics but solely evaluate "stable distributions", a similar strategy as solving equilibrium properties in statistical mechanics. This strategy is made possible by recent proof of quantum Liouville's theorem[21], a non-perturbation argument allowing the asymptotic behaviors of classical and quantum models to be treated in similar manners.

(4) Quantum Liouville's theorem indicates constant probability density $\rho$ in quantum space, just like its classical counterpart. Then the problem reduces to how to find the achievable region, for which a technique is developed, namely the *ergodic subgroup* of the evolution group formed by $\mathcal{U}^n$ [20]. By integrating over the reachable regions, one obtains the analytic $p_G$. At last, decipher its connotations for band models.

**(1) From band to Spin**. A generic two-band model $H(k, s(t))$ is expressed as

$$H(k, s(t)) = d_i(k, s(t)) \cdot \sigma_i \quad (21)$$

Phonon driving is depicted by a time-periodical parameter $s(t)$, which will distort the band and close/reopen the gap. $\sigma_i$ ($i = 1,2,3$) is Pauli matrix. $d_i$ plays the roles of magnetic field $B_i$. Equation (21) formally resembles Hamiltonian of a spin under B-field. Thus, band evolution is converted to spin's evolution under a time-dependent field. At a local $k$, we denote the two instantaneous spin eigenstates with $|n_{0,1}(t)\rangle$ (0 is ground state). Note that the instantaneous wave function $|\varphi(t)\rangle$ is likely to be different from $|n_{0,1}(t)\rangle$ as gap closing fails adiabaticity.

Since our concern is TPT, $H(k, s(t))$ should close gap (i.e., $d_i = 0$) at certain $s$. We assume gap closing happens at $k = 0$, not anywhere else. One such system is ZrTe$_5$ described by equation (4): the band cone gap is 10s meV at $\Gamma$, which is periodically closed/reopened by phonon modes, e.g., $A_{1g}$, $B_{1u}$. In ZrTe$_5$, band degeneracy at $\Gamma$ is without symmetry protection, for its point group does not have irreducible representations of dimensions > 2.

**(2) Evolution operator interpreted as rotations.** Parameterization is a key issue here. Naïvely, $\mathcal{U}$ should be expressed in the original parameters of the band model equation (4): $t$, $\omega$, $\varepsilon_0$, etc. However, the formula $\mathcal{U}(t, \omega, \varepsilon_0, ...)$ is extremely hard to find for generic cases. Even if $\mathcal{U}(t, \omega, \varepsilon_0, ...)$ is solvable in special cases, it encounters difficulty at adiabatic limit, where the dynamic phases suffer from double limit $\lim_{t\to\infty} \lim_{\omega\to 0} \int_0^t E(t') \, dt'$, whose outcome is indefinite. To avoid these issues, we adopt geometric parameters by noticing that the spin evolution operator $\mathcal{U}$ resembles $SU(2)$ rotation matrices. Consider a simple case: a spin in a magnetic field $\mathbf{B}(t)$ aligned in $z$-axis, $\mathbf{B}(t) = \frac{\Delta}{2\mu_B} \cos(\omega t) \cdot \hat{z}$,

$$\mathcal{U} = \begin{pmatrix} e^{-iE(t)} & 0 \\ 0 & e^{iE(t)} \end{pmatrix}, (22)$$

where $E(t) = -\frac{\Delta}{2}\cos(\omega t)$. In this particular case, $\mathcal{U}$ will allow the spin to stick to the original ray but only add dynamic phases. The $SU(2)$ rotation $\delta$ with fixed axis $(\alpha, \beta)$ is expressed as

$$\mathcal{R}(\alpha, \beta, \delta) = \begin{pmatrix} \cos\left(\frac{\delta}{2}\right) - i\sin\left(\frac{\delta}{2}\right)\cos(\alpha) & -i\sin\left(\frac{\delta}{2}\right)\sin(\alpha)e^{-i\beta} \\ -i\sin\left(\frac{\delta}{2}\right)\sin(\alpha)e^{i\beta} & \cos\left(\frac{\delta}{2}\right) + i\sin\left(\frac{\delta}{2}\right)\cos(\alpha) \end{pmatrix} (23)$$

We immediately recognize that dynamic evolution $\mathcal{U}$ could be re-interpreted as a rotation about $z$-axis, i.e., $\delta = \int_0^t E(t')dt' = -\frac{\Delta}{2\omega}\sin(\omega t)$, $\alpha = 0$. Therefore, by "combining" two dynamic parameters $t, \omega$ into a single rotation angle, the multiple-limit process is eluded and the divergence at adiabaticity $\omega \to 0$, $t \to \infty$ disappears. In principle, those rotation angles are functions of the original parameters, but solving $p_G$ does not refer to the particulars of $\delta(t, \omega, \varepsilon_0, ...)$, $\alpha(t, \omega, \varepsilon_0, ...)$, etc.

In equation (22), $\mathcal{U}$ is diagonal for the restriction that eigenstates are time independent. Next, we allow off-diagonals, i.e., hopping between eigenstates. Consider bending the straight path (Fig. 7a) into some angles (Fig. 7b). Fig. 7a is a special case of $\Theta = 0$ and (azimuthal angle) $\Omega = 0$.

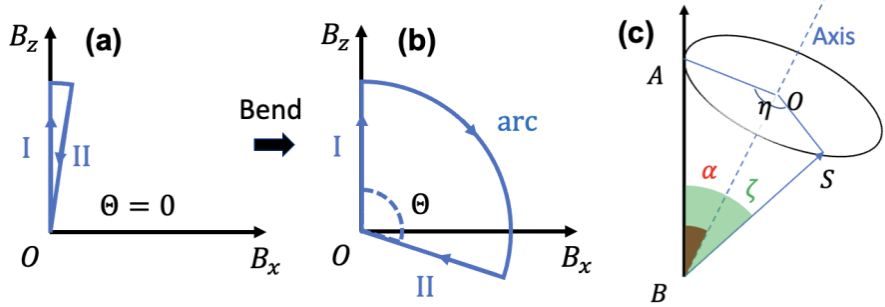

Figure. 7 (a)(b) Band parameters $d_i(t)$ are transcribed to time-dependent cyclic magnetic field $B_i(t)$, graphed with the loop. The x-z plane has azimuthal angle $\Omega = 0$. The arc section has $|\mathbf{B}| \to \infty$. (c) Spin rotates with the axis (dashed), and angle $\eta$ is in plane $OAS$.

The model is specified by a cyclic $B_i(t)$, which is denoted by the loop (blue in Fig. 7a, b). The model consists of two straight sections $I, II$ and one arc section. The $|\mathbf{B}|$ is large in the arc to ensure adiabatic evolution, the two straight sections correspond to the process of gap narrowing, and gap closing happens at the origin $O$. Previous pumping models[25,54-56] rely on adiabatic limit $\omega \ll \Delta$. This model is interesting for $\hbar\omega > \Delta_{min} = 0$.

Evidently, $\mathcal{U}$ is the product of the three sections

$$\mathcal{U} = U_{II}U_c U_I \ (24)$$

Finding $U_I$, $U_{II}$ are straightforward, and we just need to focus on the arc $U_c$. Although $\mathcal{U}$ is not adiabatic, $U_c$ is. Given a spin is along $\mathbf{B}(0)$ in arc, it is supposed to be always aligned with $\mathbf{B}(t)$. This allows one to find $U_c$'s form, as evolution's effect simply follows $\mathbf{B}$. It can be expressed with a product of two rotations

$$U_c = \mathcal{R} \cdot \Lambda. \quad (25)$$

Rotation $\mathcal{R}$ is in charge of the resultant spin orientation and will turn the spin from orientation I to II. In the particular situation of Fig. 7b, we have $\mathcal{R}\left(\frac{\pi}{2}, \Omega + \frac{\pi}{2}, \Theta\right)$, i.e., $\alpha = \frac{\pi}{2}$, $\beta = \Omega + \frac{\pi}{2}$, and $\delta = \Theta$. Additionally, there is a phase, which is adjusted by a diagonal $\Lambda$. The math origin of $U_c$ being factorizable into $\mathcal{R}$ and $\Lambda$ arises from Hopf map, i.e., $S^3 (\cong SU(2))$ is factorizable into $S^2$ sphere and a $U(1)$ phase. $\Gamma$ and $\Gamma'$ are real-valued subject to constraint $\Gamma = \Gamma'$ due to Hamiltonian's symmetry. (Sec. 4 of SI)

$$\Lambda = \begin{pmatrix} e^{-i\Gamma} & 0 \\ 0 & e^{i\Gamma'} \end{pmatrix} \quad (26)$$

We may re-write equation (25).

$$\mathcal{U} = U_{II} U_c U_I = \mathcal{R}\mathcal{R}^{-1} U_{II} \mathcal{R}\mathcal{R}^{-1} (\mathcal{R}\Lambda) U_I. \quad (27)$$

We have inserted $\mathbb{I} = \mathcal{R}\mathcal{R}^{-1}$ and plugged in $U_c = \mathcal{R}\Lambda$. Similarity transformation $\mathcal{R}^{-1} \ldots \mathcal{R}$ is equivalent to adopting eigenstates along the orientation II. With the new bases, $U'_{II} := \mathcal{R}^{-1} U_{II} \mathcal{R}$ is diagonalized. The $U_I$ is initially diagonalized.

$$U'_{II} = \begin{pmatrix} e^{-i\Phi_2} & 0 \\ 0 & e^{i\Phi_2} \end{pmatrix}, U_I = \begin{pmatrix} e^{-i\Phi_1} & 0 \\ 0 & e^{i\Phi_1} \end{pmatrix} \quad (28)$$

Then

$$\mathcal{U} = \mathcal{R} U'_{II} \Lambda U_I = \mathcal{R}\left(\frac{\pi}{2}, \Omega + \frac{\pi}{2}, \Theta\right) \begin{pmatrix} e^{-i\Phi_2} & 0 \\ 0 & e^{i\Phi_2} \end{pmatrix} \begin{pmatrix} e^{-i\Phi_c} & 0 \\ 0 & e^{i\Phi_c} \end{pmatrix} \begin{pmatrix} e^{-i\Phi_1} & 0 \\ 0 & e^{i\Phi_1} \end{pmatrix}$$

$$= \begin{pmatrix} \cos\left(\frac{\Theta}{2}\right) e^{-i\Phi} & -\sin\left(\frac{\Theta}{2}\right) e^{-i(\Omega-\Phi)} \\ \sin\left(\frac{\Theta}{2}\right) e^{i(\Omega-\Phi)} & \cos\left(\frac{\Theta}{2}\right) e^{i\Phi} \end{pmatrix}, (29)$$

where $\Phi_c = \Gamma$ and $\Phi_1 + \Phi_2 + \Phi_c = \Phi$. With $\mathcal{U}$, we may find pumping probability after $n$ cycles with equation (7).

**(3) Quantum Liouville theorem.** With evolution operator $\mathcal{U}$ equation (29), one may numerically evaluate $p_G$ with the following series (given $\Omega = 0$, i.e., in x-z plane).

$$p_1 = \sin^2\left(\frac{\Theta}{2}\right)$$

$$p_2 = \frac{1}{2}\left(p_1 + \left|-\sin\left(\frac{\Theta}{2}\right)\cos\left(\frac{\Theta}{2}\right) - e^{2i\Phi}\sin\left(\frac{\Theta}{2}\right)\cos\left(\frac{\Theta}{2}\right)\right|^2\right)$$

$$p_3 = \frac{1}{3}\left(2p_2 + \left|\cos\left(\frac{\Theta}{2}\right)\left(-\sin\left(\frac{\Theta}{2}\right)\cos\left(\frac{\Theta}{2}\right) - e^{2i\Phi}\sin\left(\frac{\Theta}{2}\right)\cos\left(\frac{\Theta}{2}\right)\right)\right.\right.$$
$$\left.\left. - \sin\left(\frac{\Theta}{2}\right)\left(-\sin^2\left(\frac{\Theta}{2}\right) + e^{-2i\Phi}\cos^2\left(\frac{\Theta}{2}\right)\right)\right|^2\right)$$
$$p_4 = \frac{1}{4}(3p_3 + \cdots) \quad (30)$$

Equation (30) is complicated, as its terms increase fast. Note that it merely gives finite orders of $p_n$ by $\frac{1}{n}\sum_j^n |(\mathcal{U}^n)_{12}|^2$ as an approximation for $p_G$. Next, we shall prove the series will eventually converge to the compact formula equation (20) at $n \to \infty$.

The key is to re-interpret the sum of the infinite series generated by $\mathcal{U}^n$ as averaging of a dynamic evolution defined by $U(t_n, 0) \coloneqq \mathcal{U}^n$, where each $\mathcal{U}$ will push the evolution forward by $\Delta t \coloneqq t_{n+1} - t_n$. Consider a point in the state space, and track its path over a long time, and find all the "footprints", i.e., the distribution $\rho$ of the "footsteps". $p_\infty$ will be determined by integration of $\rho$ over the achievable regions.

In principle, $\rho$ is not uniform, i.e., certain regions are more likely to be reached, others are less, depending on Hamiltonians. However, the long-time limit proves simple: Liouville's theorem (LT) indicates $\rho$ will approach to a constant in the achievable region. In other words, there are eventually two regions: one is unreachable and thus $\rho = 0$; the other is reachable, $\rho = \rho_0$, a constant up to normalization. Then the problem reduces to finding the reachable regions, regardless the temporal order of traversing these regions. In statistical mechanics, that is why the behaviors of a many-particle system can be handled, although it is dynamically unsolvable.

LT is independent of Hamiltonian details, beyond perturbations, suitable for treating gap-closing. However, traditional LT only works for classical physics in phase space $\{p, q\}$[29,30], it is invalid for quantum evolution, as shown by Wigner flow[57]. In recent work[21], LT is generalized to quantum, namely *quantum Liouville's theorem*.

Quantum LT makes the constant $\rho$ argument can be applied except that $\rho$ is defined in quantum space, rather than the classical phase space $\{p, q\}$[21]. Since the quantum evolution is unitary, each point in the space represents one element in a unitary group, and the achievable subspace forms a subgroup. In the case of spin, the space is SU(2) (the spin group), and the achievable space (Appx. C of Ref. [20]) is found to be a "circular orbit" (black in Fig. 7c) with axis $(\alpha, \beta)$ specified by

$$\cos^2(\alpha) = \frac{\cos^2\left(\frac{\Theta}{2}\right)\sin^2(\Phi)}{1 - \cos^2\left(\frac{\Theta}{2}\right)\cos^2(\Phi)}$$
$$\beta = \Phi - \Omega - \frac{\pi}{2} + n\pi, \quad n = 0, 1. \quad (31)$$

We define the angle between the state vector and the $+z$ axis as $\zeta$, i.e., $\Theta = \zeta$. The pumping probability depends on the projection to $(0 \quad 1)^T$ for a spin starting from $(1 \quad 0)^T$

$$\left| (0 \quad 1) \begin{pmatrix} \cos\left(\frac{\Theta}{2}\right) \\ \sin\left(\frac{\Theta}{2}\right) e^{i\Omega} \end{pmatrix} \right|^2 = \sin^2\left(\frac{\zeta}{2}\right). \quad (32)$$

We shall integrate the observable over the circular orbit weighted by the distribution density.

$$p_\infty = \frac{\int_0^{2\pi} \sin^2\left(\frac{\zeta}{2}\right) \rho(\eta) d\eta}{\int_0^{2\pi} \rho(\eta) d\eta} \quad (33)$$

$\rho(\eta)$ should be constant function as suggested by quantum LT. Then the density of points in the achievable region will be a constant $\rho_0$. The idea is in analog with classical statistical mechanics arguments: $\rho$ in an energy shell is constant. Note that equation (33) formally resembles the formulation of statistical quantity, such as

$$\langle \mathcal{O} \rangle = \frac{\sum_i \mathcal{O}_i \cdot e^{-E_i/k_B T}}{\sum_i e^{-E_i/k_B T}}, \quad (34)$$

where $\rho$ plays the role of weight factor $e^{-E_i/k_B T}$, and spin projection (pumping) is observable $\mathcal{O}$, the spin orientation $\zeta$ labeling the state corresponds to $i$ in equation (34).

From Fig. 7c, we notice the isosceles triangle $\overline{BA} = \overline{BS}$ and $\overline{OA} = \overline{OS}$. Then we use the common edge $\overline{AS}$ between triangles $BAS$ and $OAS$, to build the relation

$$\overline{BS} \cdot \sin\left(\frac{\zeta}{2}\right) = \overline{OS} \cdot \sin\left(\frac{\eta}{2}\right). \quad (35)$$

Also, we have

$$\overline{OA} = \overline{BA} \cdot \sin\left(\frac{\eta}{2}\right). \quad (36)$$

Then we can get a relation

$$\sin\left(\frac{\zeta}{2}\right) = \sin(\alpha) \sin\left(\frac{\eta}{2}\right) \quad (37)$$

Then equation (33) will become

$$p_\infty = \frac{\rho \int_0^{2\pi} \sin^2(\alpha) \sin^2\left(\frac{\eta}{2}\right) d\eta}{\rho \int_0^{2\pi} d\eta} = \sin^2(\alpha) \frac{\int_0^{2\pi} \sin^2\left(\frac{\eta}{2}\right) d\eta}{2\pi} = \frac{1}{2}\sin^2(\alpha)$$
$$= \frac{1}{2}(1 - \cos^2(\alpha)) \quad (38)$$

Then combine with equation (31), we obtain equation (20).

**(4) Project spin back to band**. The last step is to project back from the spin model to the original band model.

Based on model like equation (21), $\Theta$ will determine whether band inversion will happen at a local $k$. (Fig. 6b) If gap closing, at a certain moment, only happens at a single place (which is usually the case), the local band inversion will determine the topological state change. If smoothness for $s$ is imposed, $\Theta$ may only take two values: 0 and $\pi$, because other values, e.g., $\Theta = \frac{\pi}{2}$ will lead to an "angle" in the trajectory coordinated by $s$, violating the differentiability about $s$.

That is, smoothness of $s$ will make $\Theta$ take only two values $\Theta = 0$ and $\pi$, mapping it to a two-state topological parameter. Note that when $\Theta = 0$ and $\pi$, the value of $p_G$ becomes independent of $\Phi$. Even with $\Theta = 0, \Phi = 0$, $p_G$ encounters "0/0", the nominator approaches faster, it still converges to 0. As such, one plugs $\Theta = 0$ and $\pi$ into equation (20), obtaining $p_G = 0$ or $\frac{1}{2}$ depending on whether TPT takes place.

Note that at $\Theta = 0$ and $\pi$, the value of equation (20) tends to be independent of dynamic phase $\Phi$. This confirms our numerical finding in Fig. 2 that given TPT, adjusting the energetic parameter $\varepsilon_0$ only leads to a plateau. This confirmation can only be made with analytic $p_G$; for finite-order $p_n$ (equation (30)), the expression always mixes the dependence of $\Phi$ and $\Theta$.

3. **Simulation of pumping curves & Time dependent ensemble.**

Our second goal is to derive equation (8). A statistical observable is yielded by Ref. [58]

$$\mathcal{O}(t) = \text{tr}[\hat{O}\hat{\rho}_t] \quad (39)$$

In this case, observable $\mathcal{O}$ stands for the pumping probability of a two-level system from $|n_0\rangle$ to $|n_1\rangle$, the corresponding operator reads

$$\hat{O} = \begin{pmatrix} 0 & 0 \\ 0 & 1 \end{pmatrix} \quad (40)$$

Plug in, we have

$$\mathcal{O}(t) = \langle n_1|\hat{\rho}_t|n_1\rangle$$
$$= \frac{1}{n}(\langle n_1|U(t,0)\hat{\rho}_0 U(0,t)|n_1\rangle + \langle n_1|U(t+\Delta t,0)\hat{\rho}_0 U(0,t+\Delta t)|n_1\rangle$$
$$+ \cdots \langle n_1|U(t+(n-1)\Delta t,0)\hat{\rho}_0 U(0,t+(n-1)\Delta t)|n_1\rangle) \quad (41)$$

We start with the ground state,

$$\hat{\rho}_0 = |\varphi_{t=0}\rangle\langle\varphi_{t=0}| \quad (42)$$

Plug in, equation (41) becomes

$$p_\varepsilon(t) = \frac{1}{n}(\langle n_1|U(t,0)|\varphi_{t=0}\rangle\langle\varphi_{t=0}|U(0,t)|n_1\rangle + \cdots$$
$$+ \langle n_1|U(t+(n-1)\Delta t,0)|\varphi_{t=0}\rangle\langle\varphi_{t=0}|U(0,t+(n-1)\Delta t)|n_1\rangle)$$
$$= \frac{1}{n}(|\langle n_1|U(t,0)|\varphi_{t=0}\rangle|^2 + \cdots + |\langle n_1|U(t+(n-1)\Delta t,0)|\varphi_{t=0}\rangle|^2)$$
$$= \frac{1}{n}\sum_j^n p_j(t) \quad (43)$$

where

$$p_j(t) = |\langle n_1|U(t+(j-1)\Delta t,0)|\varphi_{t=0}\rangle|^2 \quad (44)$$

In this particular case $p_j(t)$ stands for the $j^{th}$ system's pumping probability at $t$. We stipulate the $j=1$ is the earliest small region in the light spot to start vibration, the $j=n$ is the latest. Then, we have the following relation,

$$p_{j+1}(t) = p_j(t-\Delta t) \quad (45)$$

More generally

$$p_{j+l}(t) = p_j(t-l\Delta t) \quad (46)$$

Then equation (43) becomes

$$p_\varepsilon(t) = \frac{1}{n}\sum_j^n p_1(t-j\Delta t) \quad (47)$$

Then $p_1(t)$ could be evaluated by $\mathcal{U}^n(\Theta,\Phi,\Omega)$ with $\Theta = \pi$. Because $\Theta = \pi$ corresponds to the gap closing will lead to TPT, Resulting in the form like Fig. 4c.

Summary of parameters involved in simulating GP. (i) $\tau_{measure}$ is averaging time scale for an observable. For ultra-fast spectrum, measurement preciseness is $\tau_{measure} \lesssim 50\ fs$. We just take $\tau_{measure} \to 0$, since this time scale is much smaller than any others. (ii) $\tau_{cycle}$ means the time interval between two times of TPT. If we assume each phonon cycle closes up the gap once, $\tau_{cycle}$ is the same as phonon's period. The typical phonon has $\tau_{cycle} \sim 1\ ps$. In this case, we adopt $\tau_{cycle} = 0.83\ ps$ to describe $A_{1g}$ phonon in ZrTe$_5$. (iii) $\tau_{ergodic}$ physically means the time required by the area in the light spot start all starting phonon vibration. It will determine (but *not* straightforwardly equal to) the time of reaching the plateau in Fig. 4c. Here, we adopt $\tau_{ergodic} \sim 6\ ps$ from experiment. Thus, simulation in Fig. 4c can well capture the trend, while the time scales are not its predicting power, since it relies on details of the sample and interaction between light and material. (iv) $\Delta t$ is the evolution time mismatch for different systems in the ensemble. $\Delta t$ should be small compared with $\tau_{cycle}$; in this case, we adopt $\Delta t = 0.1\ ps$.

**Competing interests.** The authors declare that they have no competing interests.

**Data and materials availability.** All data needed to evaluate the conclusions in the paper are present in the paper. Additional data related to this paper may be available on reasonable request from B. Q. S. and J.W.

**References.**